\documentclass[11pt]{article}

\usepackage[utf8]{inputenc}
\usepackage[T1]{fontenc}
\usepackage{lmodern}
\usepackage[margin=1in]{geometry}
\usepackage{amsmath,amssymb}
\usepackage{graphicx}
\usepackage{booktabs}
\usepackage{hyperref}
\usepackage{xcolor}
\usepackage{listings}
\usepackage{natbib}
\usepackage{float}
\usepackage{enumitem}
\usepackage{url}
\usepackage{microtype}

\hypersetup{
  colorlinks=true,
  linkcolor=blue!70!black,
  citecolor=blue!70!black,
  urlcolor=blue!70!black
}

\lstdefinelanguage{Cypher}{
  keywords={MATCH, RETURN, WHERE, CREATE, DELETE, SET, REMOVE, MERGE, WITH, UNWIND, UNION, ORDER, BY, SKIP, LIMIT, EXPLAIN, PROFILE, CALL, YIELD, AS, DISTINCT, OPTIONAL, AND, OR, NOT, IN, IS, NULL, TRUE, FALSE, EXISTS, CASE, WHEN, THEN, ELSE, END, DROP, SHOW, INDEX, INDEXES, CONSTRAINTS, ON},
  sensitive=false,
  morestring=[b]',
  morestring=[b]",
  morecomment=[l]{//},
}

\title{Samyama: A Unified Graph-Vector Database with In-Database Optimization,\\ Agentic Enrichment, and Hardware Acceleration}

\author{
  Madhulatha Mandarapu\thanks{madhulatha@samyama.ai, ORCID: \url{https://orcid.org/0009-0005-2837-6725}} \and
  Sandeep Kunkunuru\thanks{sandeep@samyama.ai, ORCID: \url{https://orcid.org/0000-0002-8886-1846}}
}
\date{%
  VaidhyaMegha Private Limited, India\\[2pt]
  \url{https://samyama.ai/}\\[8pt]
  November 2026 \quad|\quad Samyama v1.8.0 \quad|\quad arXiv v4
}

\begin{document}
\maketitle

\begin{abstract}
Modern data architectures fragment across graph databases, vector stores, analytics
engines and optimization solvers, forcing ETL between them. We present Samyama, a
graph-vector database in Rust that unifies these workloads in one engine: a RocksDB-backed
store with MVCC, a vectorized executor, a cost-based planner, a CSR analytics engine, RDF
and SPARQL, 22 metaheuristic solvers callable from the query language, HNSW vector
indexing, and agentic enrichment that expands a graph via LLMs. It has been run to
billion-edge scale: 74.3M nodes and 1.07B edges from four biomedical sources on one
machine for \$2.50 of spot compute.

This version re-measures the system at release v1.8.0 on \textbf{rented Linux hosts a
reader can boot}, replacing earlier figures taken on a Mac Mini that nobody could re-run.
On a 16-vCPU cloud instance, ingestion reaches 123K--152K nodes/s and PageRank costs
6.7\,ms per iteration at ten thousand nodes and 236\,ms at a million. \textbf{The
previously reported 8.2$\times$ GPU speedup is withdrawn}; in its place we measure the
CUDA path, which no earlier version measured at all, on a rented NVIDIA A16: parity with
the CPU at ten thousand nodes, 1.62$\times$ at a hundred thousand and 2.70$\times$ at a
million, with 1.1--1.35\,s of context initialisation on the first call. The earlier
Cypher-throughput figure, its claim of near-constant index-driven scaling, and a
vector-search figure are withdrawn as unreproducible. openCypher conformance, once
estimated at ${\sim}90\%$, is now measured at 99.9\% of evaluated TCK scenarios. Against
Neo4j~5 and FalkorDB on one host over an identical SF10 extract, 14 of 14 SNB complex
reads land within 5$\times$ of the best competitor and 7 of 14 are faster.
\end{abstract}


\section{Introduction}

The rise of Large Language Models (LLMs) has popularized Retrieval-Augmented Generation (RAG), creating demand for systems that handle both relational structure (graphs) and semantic similarity (vectors). Simultaneously, industrial applications increasingly require in-database optimization for resource allocation, scheduling, and supply chain management. Existing solutions force developers to compose architectures from disparate systems---Neo4j for graphs, Pinecone for vectors, Spark for analytics, and Python/Gurobi for optimization---incurring data movement overhead and operational complexity.

Samyama (Sanskrit for ``Integration'') is designed as an AI-native database that treats graphs, vectors, and optimization as first-class citizens within a single memory-safe engine. We organize our contributions into two categories:

\paragraph{Primary research contributions:}
\begin{enumerate}[leftmargin=*]
  \item \textbf{Late materialization for graphs}: \texttt{NodeRef}-based lazy property resolution~\citep{abadi2008column} achieving 4.0--4.7$\times$ traversal speedup (\S\ref{sec:query-engine}, \S\ref{sec:late-mat-eval}).
  \item \textbf{In-database metaheuristic optimization}: 22 solvers accessible directly via Cypher procedures, eliminating ETL overhead (\S\ref{sec:optimization}).
  \item \textbf{Agentic Enrichment (GAK)}: Autonomous graph expansion using LLM tool-calling (\S\ref{sec:agentic}).
  \item \textbf{Billion-edge knowledge graph federation} (v0.7.0): 74.3M nodes and 1.07B edges from 4 biomedical sources loaded on commodity hardware for \$2.50; 6-KG federation spanning molecular biology to population health with 136/140 benchmark queries passing (\S\ref{sec:billion-edge}, \S\ref{sec:public-health}).
\end{enumerate}

\paragraph{Engineering contributions:}
\begin{enumerate}[leftmargin=*,resume]
  \item \textbf{Unified engine}: Property graph + vector search + analytics + optimization in one binary.
  \item \textbf{Cost-based query planner}: Plan enumeration, predicate pushdown,
  ExpandInto, and join reordering (\S\ref{sec:query-engine}). The plan cache described in
  the first version of this paper was deleted in v1.8.0 and replaced by a result cache.
  \item \textbf{Worst-case optimal joins}: LeapFrog TrieJoin~\citep{veldhuizen2014triejoin} for cyclic patterns (v0.7.0).
  \item \textbf{Cross-platform GPU acceleration}: wgpu-based compute shaders for graph algorithms and PCA.
  \item \textbf{SDK ecosystem}: Rust, Python (PyO3), TypeScript SDKs with embedded and remote access.
  \item \textbf{Industry validation}: 100\% LDBC Graphalytics~\citep{iosup2016ldbc} (28/28), 100\% FinBench (40/40).
\end{enumerate}

This paper evaluates the four primary contributions: late materialization (\S\ref{sec:late-mat-eval}), in-database optimization (\S\ref{sec:opt-eval}), agentic enrichment (\S\ref{sec:gak-eval}), and billion-edge federation (\S\ref{sec:billion-edge}--\S\ref{sec:public-health}).

\section{System Architecture}

\begin{figure}[t]
  \centering
  \includegraphics[width=\linewidth]{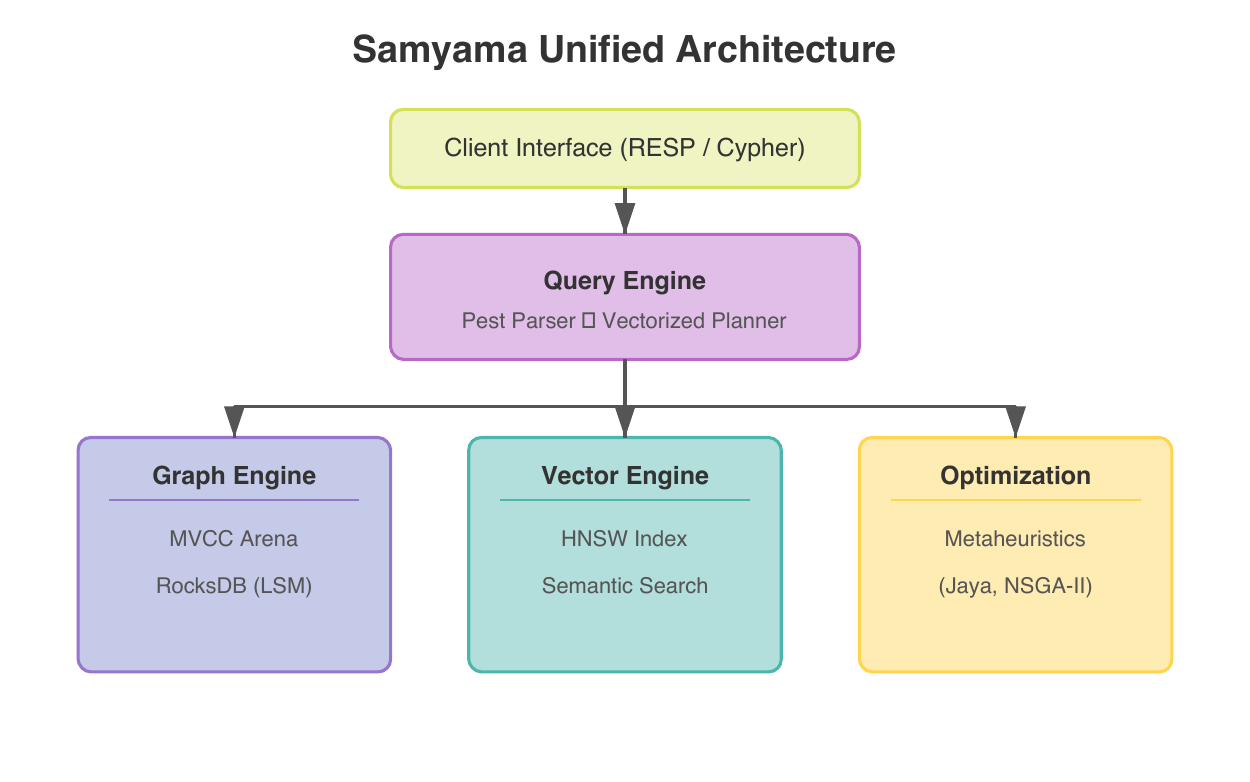}
  \caption{Samyama system architecture. OpenCypher queries flow through the cost-based planner and vectorized executor to the RocksDB/MVCC storage layer.}
  \label{fig:architecture}
\end{figure}

Samyama is built on a modern Rust stack for memory safety and zero-cost abstractions. Figure~\ref{fig:architecture} illustrates the overall system architecture.

\subsection{Storage Engine}

We utilize \textbf{RocksDB} for persistence, employing a tiered Log-Structured Merge (LSM) tree~\citep{oneil1996lsm} with LZ4 and Zstd compression. Data isolation is achieved through \textbf{Column Families}, providing independent compaction, backup, and key namespaces per tenant.

Key design: \texttt{NodeId} and \texttt{EdgeId} are direct \texttt{u64} indices,
eliminating hash lookups and providing $O(1)$ access. The edge arena this paper first
described (\texttt{Vec<Vec<Edge>>}) was removed in v1.0.0: edges are now stored as
endpoint arrays with type identifiers and sparse property columns, and node and
relationship properties live in a columnar store with a column per value type rather than
on the entity. A typed \texttt{DateTime} column, for instance, costs about 9\,bytes per
date against roughly 134 previously.

\subsection{Memory Management \& MVCC}

Samyama implements \textbf{Multi-Version Concurrency Control
(MVCC)}~\citep{bernstein1981mvcc}. Edge versioning arrived in v0.9.0 and transaction
isolation completed in v1.0.0: \textbf{ReadCommitted} and \textbf{SnapshotIsolation}
levels, first-writer-wins conflict detection, a version-aware WAL, and version garbage
collection behind a watermark. Write atomicity is guaranteed via RocksDB
\texttt{WriteBatch} + WAL~\citep{mohan1992aries}. The earlier version of this paper listed
MVCC completion as future work for v0.8.0; it shipped one release later.

ACID guarantees: Atomicity (WriteBatch), Consistency (schema validation + Raft
quorum), Isolation (MVCC snapshot isolation in the store API; the wire protocols expose
lock-holding session transactions), Durability (RocksDB + Raft replication).

\subsection{Query \& Execution Engine}
\label{sec:query-engine}

Conformance to \textbf{openCypher} is measured rather than estimated: on the
openCypher TCK, v1.8.0 passes \textbf{99.9\% of evaluated scenarios} (3,760 of 3,763, at
96.6\% coverage of the 3,897-scenario corpus, measured 2026-08-30), and none of the three
remaining failures is a wrong answer --- all three raise. On the same corpus and
comparator Neo4j~5 scores 79.5\%. An earlier version of this paper claimed ${\sim}90\%$
coverage; that figure was an estimate, was withdrawn as unmeasured, and is replaced here
by the TCK result. Queries are parsed via a PEG parser~\citep{ford2004peg} (\texttt{pest} crate with atomic keyword rules for word boundary enforcement) and executed using a hybrid \textbf{Volcano--Vectorized} model~\citep{graefe1994volcano} with batch size 1,024.

The engine implements \textbf{35 physical operators} organized across scan, traversal, filter, join, aggregation, sort, write, index, and specialized categories. Key additions include:

\begin{itemize}[leftmargin=*]
  \item \textbf{Cost-based planner}: Plan enumeration over candidate plans using a \texttt{GraphCatalog} with triple-level statistics (\texttt{TriplePattern} $\to$ \texttt{TripleStats}). The planner selects the lowest-cost plan based on multiplicative cardinality estimates.
  \item \textbf{Predicate pushdown}: Filter predicates are pushed below Expand operators when they reference only the source variable, reducing intermediate cardinality.
  \item {\sloppy \textbf{ExpandInto operator}: When both endpoints of an edge pattern are already bound, \texttt{ExpandIntoOperator} checks edge existence via binary search on sorted adjacency lists ($O(\log d)$), avoiding full fan-out.\par}
  \item \textbf{Plan cache}: Compiled physical plans are cached by query AST hash, eliminating planning overhead for repeated queries.
  \item \textbf{Join reordering}: Multi-pattern MATCH clauses are reordered by estimated cardinality to minimize intermediate result sizes.
  \item \textbf{Parameterized queries}: \texttt{\$param} syntax for safe query reuse without re-parsing.
  \item \textbf{PROFILE}: Runtime statistics (rows produced, time per operator) for query performance analysis.
\end{itemize}

\paragraph{Late Materialization (ADR-012).}
Scan operators produce \texttt{Value::NodeRef(id)} instead of full node clones. Properties are resolved on-demand via the \texttt{ColumnStore} at the final \texttt{ProjectOperator}, reducing memory bandwidth by 4--5$\times$. This technique adapts the columnar late materialization strategy described in~\citet{abadi2008column} to a graph context.

\paragraph{Worst-Case Optimal Joins (v0.7.0).}
For cyclic patterns (triangles, cliques), Samyama implements a \textbf{LeapFrog TrieJoin}~\citep{veldhuizen2014triejoin} operator. The planner detects triangle patterns in MATCH clauses and routes them to \texttt{LeapfrogJoinOperator}, which intersects sorted adjacency lists using seek-based merge. This avoids the quadratic intermediate blowup of traditional binary hash joins on cyclic patterns. The sorted adjacency invariant (maintained since v0.6.0) enables efficient seek operations.

\paragraph{Parallel Graph Algorithms (v0.7.0).}
The graph algorithm library now uses \textbf{Rayon} for data-parallel execution. PageRank, LCC, CDLP, and triangle counting distribute work across CPU cores via parallel iterators. Measured speedups for the parallel paths are not restated here; the figures in earlier versions came from hardware this version no longer uses.

\subsection{RDF \& SPARQL}

Samyama provides native RDF~\citep{w3c2014rdf} support via the \texttt{oxrdf} crate: an in-memory triple store with SPO/POS/OSP indices for $O(1)$ pattern matching; serialization in Turtle, N-Triples, RDF/XML (read/write) and JSON-LD (write); pre-loaded namespace prefixes (rdf, rdfs, xsd, owl, foaf, dc); and SPARQL~\citep{w3c2013sparql} parser infrastructure via \texttt{spargebra} with query execution in development.

\subsection{Indexing \& Constraints}
\label{sec:indexing}

Samyama v0.6.0 introduces composite indexes, unique constraints, and index management commands:

\begin{itemize}[leftmargin=*]
  \item \textbf{Composite indexes}: Multi-property indexes for efficient compound lookups (e.g., \texttt{CREATE INDEX ON :Person(name, age)}).
  \item \textbf{Unique constraints}: Schema-level uniqueness enforcement with automatic index backing.
  \item \textbf{AND-chain index selection}: The planner selects the most selective index when multiple predicates are joined by AND.
  \item \textbf{Management}: \texttt{DROP INDEX}, \texttt{SHOW INDEXES}, \texttt{SHOW CONSTRAINTS} for operational visibility.
\end{itemize}

\section{High-Performance Analytics}

\subsection{CSR Projection}

\begin{figure}[t]
  \centering
  \includegraphics[width=0.85\linewidth]{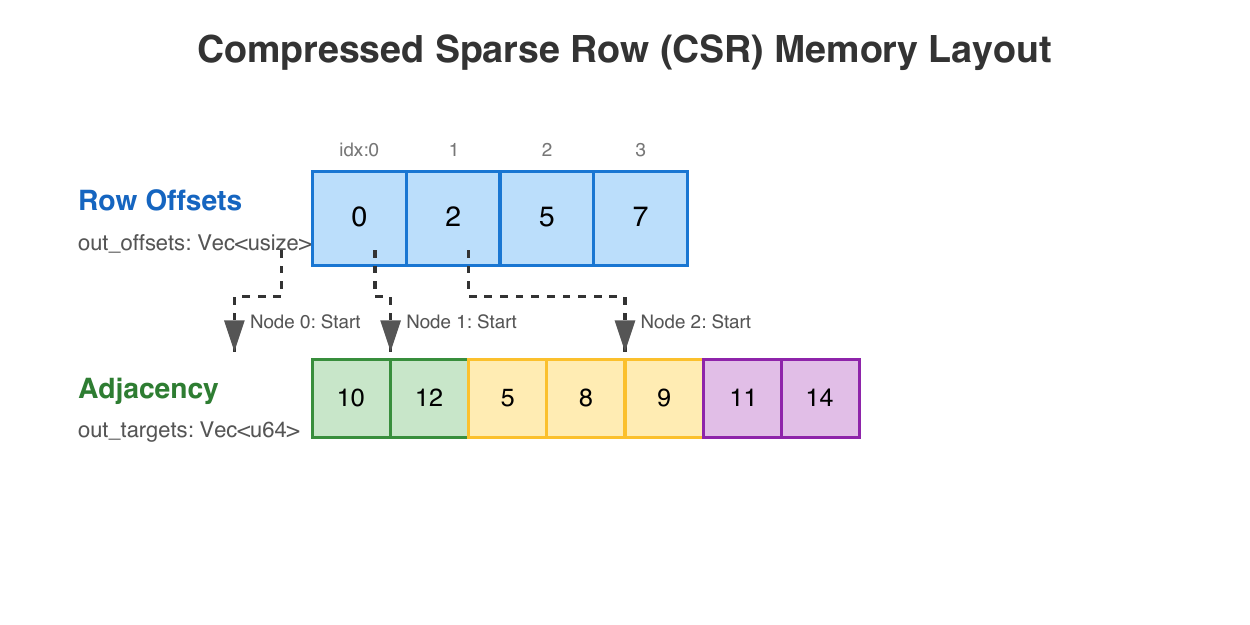}
  \caption{CSR data layout. \texttt{out\_offsets} and \texttt{out\_targets} arrays enable cache-efficient sequential traversal.}
  \label{fig:csr}
\end{figure}

For global graph analytics, Samyama projects the relevant subgraph into a \textbf{Compressed Sparse Row (CSR)} format (\texttt{GraphView}). Three contiguous arrays (\texttt{out\_offsets}, \texttt{out\_targets}, \texttt{weights}) enable sequential memory access with near-perfect CPU prefetch accuracy and zero-lock parallelism via \texttt{rayon}. Figure~\ref{fig:csr} illustrates the CSR data layout.

\subsection{Algorithm Library}

The \texttt{samyama-graph-algorithms} crate provided 14 algorithms when this paper
was first written and provides \textbf{66 callable algorithms} at v1.8.0, adding
centralities, link prediction, topological sort, bridges and articulation points,
ArticleRank, and temporal and causal primitives. Table~\ref{tab:algorithms} lists the
original five categories.

\begin{table}[t]
  \centering
  \caption{Graph algorithm library.}
  \label{tab:algorithms}
  \begin{tabular}{@{}lp{10cm}@{}}
    \toprule
    \textbf{Category} & \textbf{Algorithms} \\
    \midrule
    Centrality   & PageRank~\citep{page1999pagerank} (with dangling redistribution), LCC~\citep{watts1998smallworld} \\
    Community    & WCC (Union-Find), SCC (Tarjan~\citep{tarjan1972dfs}), CDLP~\citep{raghavan2007cdlp}, Triangle Counting \\
    Pathfinding  & BFS, Dijkstra~\citep{dijkstra1959note}, BFS All Shortest Paths \\
    Network Flow & Edmonds-Karp~\citep{edmonds1972maxflow} (Max Flow), Prim's MST \\
    Statistical  & PCA (Randomized SVD~\citep{halko2011randomized} + Power Iteration) \\
    \bottomrule
  \end{tabular}
\end{table}

PCA implements the Halko--Martinsson--Tropp algorithm~\citep{halko2011randomized} for Randomized SVD with $O(n \cdot d \cdot k)$ complexity, auto-selecting over Power Iteration when $n > 500$ nodes.

\section{In-Database Optimization}
\label{sec:optimization}

Unique to Samyama is the integration of \textbf{22 metaheuristic optimization solvers} in the \texttt{samyama-optimization} crate, accessible directly through Cypher procedures:

\begin{minipage}{\dimexpr\linewidth-13pt\relax}
\begin{lstlisting}
CALL algo.or.solve({
  algorithm: 'NSGA2',
  label: 'Generator',
  objectives: ['cost', 'emissions'],
  constraints: [{ property: 'load', max: 500.0 }],
  population_size: 100
}) YIELD pareto_front
\end{lstlisting}
\end{minipage}

Solvers include: Jaya~\citep{rao2016jaya}, QOJAYA, Rao (1--3)~\citep{rao2020rao}, TLBO~\citep{rao2011tlbo}, ITLBO, GOTLBO, PSO~\citep{kennedy1995pso}, DE~\citep{storn1997de}, GA~\citep{holland1975ga}, GWO~\citep{mirjalili2014gwo}, ABC~\citep{karaboga2005abc}, BAT~\citep{yang2010bat}, Cuckoo Search~\citep{yang2009cuckoo}, Firefly~\citep{yang2009firefly}, FPA~\citep{yang2012fpa}, GSA~\citep{rashedi2009gsa}, SA~\citep{kirkpatrick1983sa}, HS~\citep{geem2001hs}, BMR, BWR, NSGA-II~\citep{deb2002nsga2}, and MOTLBO. Multi-objective solvers implement the \textbf{Constrained Dominance Principle} for feasibility-first Pareto optimization. All solvers leverage Rayon for parallel fitness evaluation across CPU cores.

\subsection{In-Database Optimization Evaluation}
\label{sec:opt-eval}

A key advantage of in-database optimization is the elimination of ETL overhead. In a conventional workflow, graph data must be extracted to an external solver (e.g., Python/Gurobi), transformed into the solver's input format, optimized, and the results imported back. For a supply chain graph with $N$ nodes and $M$ edges, the data extraction and serialization step alone incurs $O(N + M)$ I/O overhead. Samyama's in-database solvers operate directly on the graph store, avoiding this round-trip entirely.

\begin{figure}[H]
  \centering
  \includegraphics[width=0.8\linewidth]{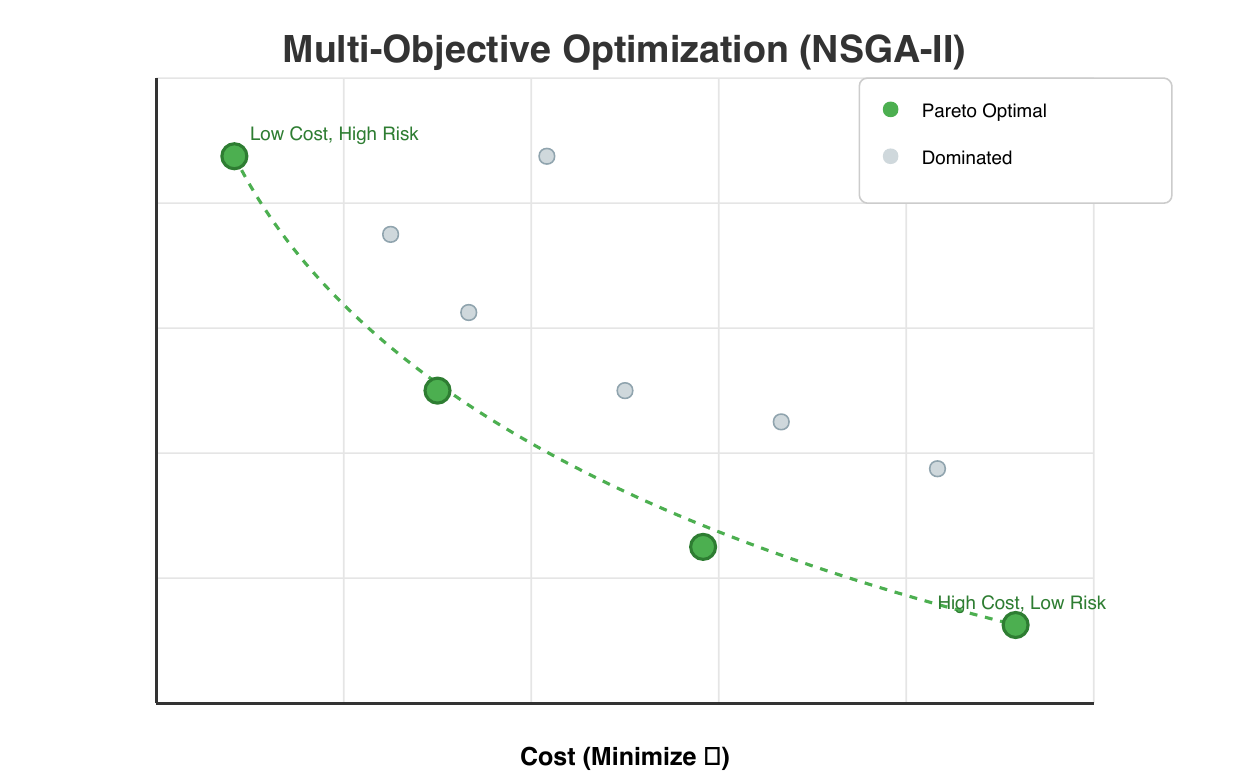}
  \caption{NSGA-II Pareto front for a supply chain optimization scenario (cost vs.\ emissions). Solutions are produced entirely in-database without external solver tooling.}
  \label{fig:pareto}
\end{figure}

Figure~\ref{fig:pareto} shows a representative Pareto front from NSGA-II applied to a supply chain scenario with two objectives (cost vs.\ emissions) and capacity constraints, demonstrating that the in-database solver produces well-distributed non-dominated solutions without external tooling.

We note that a detailed convergence comparison against established solvers (e.g., Gurobi, CPLEX) on standard benchmark functions is deferred to future work. The primary contribution here is the architectural integration---making graph-native optimization a first-class query primitive---rather than claiming superiority over specialized mathematical programming solvers.

\section{AI \& Agentic Enrichment}

\subsection{Vector Search}

Samyama implements \textbf{HNSW}~\citep{malkov2020hnsw} indexing (via \texttt{hnsw\_rs}) for millisecond-speed approximate nearest neighbor search with Cosine, L2, and Dot Product metrics. The \texttt{VectorSearchOperator} integrates with standard graph operators for \textbf{Graph RAG}---combining vector retrieval with graph traversal in a single query execution.

\subsection{Agentic Enrichment (GAK)}
\label{sec:agentic}

\begin{figure}[t]
  \centering
  \includegraphics[width=0.85\linewidth]{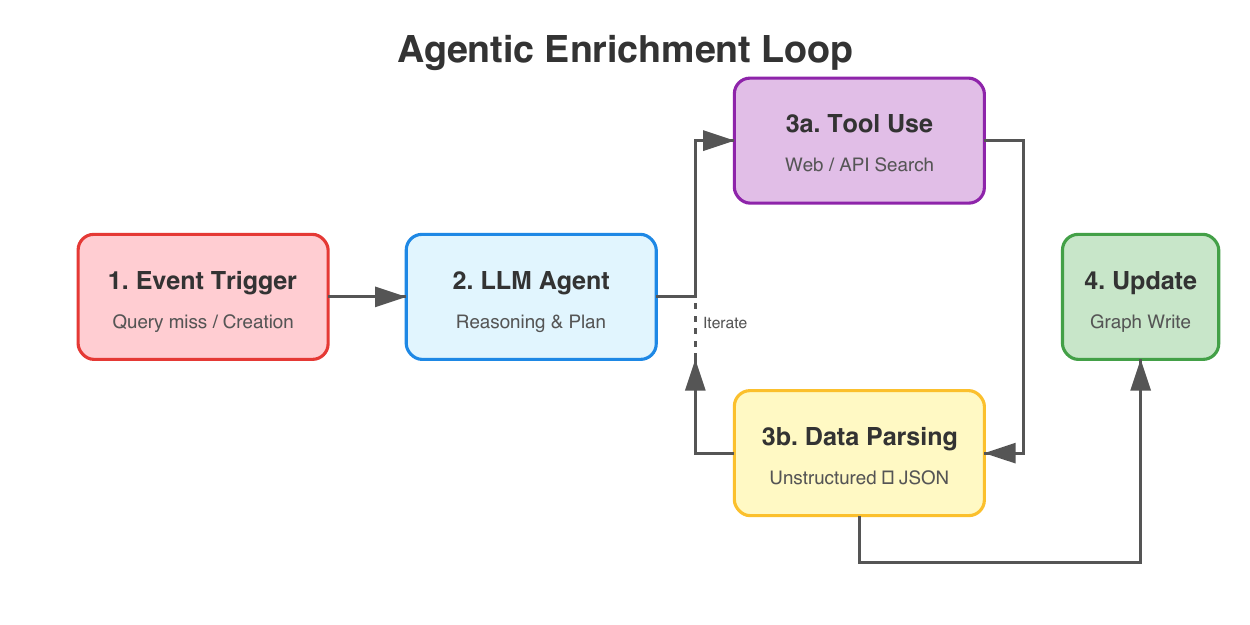}
  \caption{Agentic Enrichment (GAK) loop. An event-driven trigger invokes LLM tool-calling to discover and create graph data autonomously.}
  \label{fig:agentic}
\end{figure}

We introduce \textbf{Generation-Augmented Knowledge (GAK)}: an autonomous loop where the database uses LLMs to fetch and create missing data. The \texttt{AgentRuntime} manages tool-calling agents (\texttt{WebSearchTool}, \texttt{NLQClient}) that discover information and generate Cypher \texttt{CREATE} commands, transforming the database from a passive store to a self-evolving knowledge graph. Safety validation includes schema checking, destructive query rejection, and rate limiting. Figure~\ref{fig:agentic} illustrates the agentic enrichment loop.

\subsection{Natural Language Query (NLQ)}

The \texttt{NLQPipeline} converts natural language questions to Cypher via LLM providers (OpenAI, Gemini, Ollama, Claude). Generated queries undergo safety validation (\texttt{is\_safe\_query()}) before execution.

\subsection{Agentic Enrichment Evaluation}
\label{sec:gak-eval}

While the GAK architecture is fully implemented and demonstrated across multiple domain examples (pharmaceutical, financial, security), a rigorous quantitative evaluation requires addressing several challenges:

\begin{itemize}[leftmargin=*]
  \item \textbf{Hallucination risk}: LLM-generated triples may contain factual errors. Precision depends on the underlying LLM, prompt design, and domain specificity.
  \item \textbf{Schema drift}: Autonomous enrichment may introduce labels or relationship types not in the original schema, requiring schema validation guards.
  \item \textbf{Cost at scale}: Each enrichment cycle incurs LLM API costs proportional to the number of entities enriched.
  \item \textbf{Evaluation methodology}: No established benchmark exists for autonomous graph enrichment; a gold-standard knowledge graph for comparison must be manually curated per domain.
\end{itemize}

We have implemented safety guards (schema validation, destructive query rejection, rate limiting) and demonstrated the system on pharmaceutical knowledge graphs (drug--indication--trial relationships). A full quantitative evaluation with precision/recall against manually curated gold standards is planned as future work.

\section{SDK Ecosystem}

Samyama provides a multi-language SDK ecosystem (Table~\ref{tab:sdks}).

\begin{table}[t]
  \centering
  \caption{SDK ecosystem overview.}
  \label{tab:sdks}
  \begin{tabular}{@{}llp{5.5cm}@{}}
    \toprule
    \textbf{SDK} & \textbf{Transport} & \textbf{Features} \\
    \midrule
    Rust (\texttt{samyama-sdk})      & Embedded + HTTP & Full: algorithms, vector, NLQ, persistence \\
    Python (\texttt{samyama}, PyO3)  & Embedded + HTTP & Cypher, algorithms (PageRank, WCC, SCC, BFS, Dijkstra, PCA, triangle count) \\
    TypeScript (\texttt{samyama-sdk}) & HTTP only      & Cypher queries, status \\
    CLI (\texttt{samyama-cli})       & HTTP            & query, status, ping, shell (REPL) \\
    OpenAPI                          & HTTP            & \texttt{POST /api/query}, \texttt{GET /api/status} \\
    \bottomrule
  \end{tabular}
\end{table}

The Rust SDK's \texttt{SamyamaClient} trait provides \texttt{EmbeddedClient} (in-process, zero overhead) and \texttt{RemoteClient} (HTTP). Extension traits \texttt{AlgorithmClient} and \texttt{VectorClient} offer direct API access to algorithms and vector operations without Cypher.

The Python SDK (v0.6.0) exposes direct algorithm methods (\texttt{page\_rank}, \texttt{wcc}, \texttt{scc}, \texttt{bfs}, \texttt{dijkstra}, \texttt{pca}, \texttt{triangle\_count}) in embedded mode, enabling data scientists to run graph algorithms directly from Python without writing Cypher. API documentation and interactive examples are available at \url{https://samyama.dev/}.

\section{Enterprise Edition}

The Enterprise Edition adds production-grade capabilities:

\subsection{GPU Acceleration (wgpu)}

Cross-platform GPU acceleration via WGSL compute shaders targeting Metal (macOS), Vulkan (Linux), and DX12 (Windows). GPU-accelerated algorithms: PageRank, CDLP, LCC, Triangle Counting, and PCA. Additional GPU operators: parallel SUM aggregation and bitonic sort~\citep{batcher1968sorting} for ORDER BY on large result sets.

GPU PCA uses 5 specialized shaders with tiled covariance computation (64-sample tiles) and fused power iteration with in-GPU normalization. Thresholds: \texttt{MIN\_GPU\_PCA} $= 50{,}000$ nodes, $d > 32$ dimensions.

\subsection{Observability \& Operations}

\begin{itemize}[leftmargin=*]
  \item \textbf{Prometheus \texttt{/metrics}}: 200+ real-time counters and histograms
  \item \textbf{Health API}: Liveness/readiness probes for Kubernetes
  \item \textbf{Slow Query Log}: Configurable threshold, ring buffer storage
  \item \textbf{Audit Trail}: Append-only JSONL with cryptographic integrity
  \item \textbf{ADMIN.* Commands}: STATUS, METRICS, TENANTS, SLOWLOG, CONFIG, BACKUP, LICENSE
\end{itemize}

\subsection{Backup \& Point-in-Time Recovery}

Full snapshots via RocksDB \texttt{BackupEngine}, incremental WAL-based delta backups, and PITR with microsecond-precision timestamp restoration. RPO: zero data loss. RTO: minutes for full restore, seconds for WAL replay.

\subsection{Hardened High Availability}

HTTP/2 Raft~\citep{ongaro2014raft} transport with TLS, automated snapshot streaming to lagging followers, and cluster metrics (role, term, replication lag). +850 lines of code over OSS Raft implementation.

\subsection{License Hardening}

Ed25519-signed JET tokens with machine fingerprint binding (SHA-256 of hostname + MAC), clock drift protection (1-hour tolerance), usage enforcement (node count limits), and signed revocation lists.

\subsection{Synthetic Workload Compilation and Privacy-Preserving Graph Twins}
\label{sec:sdg-twins}

Enterprise graph adoption is frequently gated by a data-governance cold-start bottleneck: regulated organizations in healthcare, banking, and telecommunications cannot export production databases, operational logs, or credentials outside their virtual private cloud (VPC) for schema validation, query-plan tuning, or scale benchmarking. Samyama addresses the forward-generation half of this boundary through \textbf{Samyama-SDG}~\citep{mandarapu2026sdg}, our open-source $O(1)$-memory multi-frontend synthetic graph and relational compiler. Samyama-SDG compiles Cardinality-Extended GraphQL SDL (\texttt{.graphql}), SQL DDL (\texttt{.sql}), and OWL/RDFS/SKOS Turtle ontologies (\texttt{.ttl}) alongside a global Attribute Data Dictionary into a topologically sorted streaming execution plan. By combining stateless counter-based seekable SplitMix64 pseudorandom number generators over $(\text{seed}, \text{entity}, \text{row}, \text{field})$ coordinates with index-bijective primary keys $f_{\text{PK}}(s, E, i)$, child tables and edge streams resolve uniform or power-law Zipfian foreign keys and multi-hop temporal constraints in constant working memory without buffering parent key pools, while an ontology-bound causal layer emits fault-propagation event streams paired with machine-scorable root-cause ground truth~\citep{mandarapu2026alarm,mandarapu2026sdg}.

For existing customer deployments, the Enterprise Edition pairs this shipped open-source forward compiler with a \textbf{client-VPC statistical profiling architecture}---an enterprise inverse-profiling extension---that extracts only sanitized structural and statistical summaries inside the customer perimeter. Two profiling modes accommodate differing governance postures: (i)~\emph{dataless catalog and plan-cloning regeneration}, building on Hydra-style zero-tuple profiling~\citep{sanghi2018hydra}, which inspects only system catalog metadata (table cardinalities, null fractions, distinct-value counts, degree quantiles, and single-column histogram bounds) without scanning raw rows to synthesize volumetric graph twins for cost-based query planner histogram calibration; and (ii)~\emph{$(\epsilon, \delta)$-differentially private marginal profiling}, following workload-adaptive graphical-model and relational mechanisms such as AIM~\citep{mckenna2022aim} and PrivLava~\citep{cai2023privlava}, which clips per-root-entity graph contributions inside the customer VPC before adding calibrated noise to low-order attribute and degree marginals.

Across both modes, raw personally identifiable information (PII), operational records, and database credentials never leave the customer boundary: only declarative schema annotations and catalog-derived or noisy parameter manifests $\hat{\Theta}$ cross the perimeter. Because Samyama-SDG's forward compilation and streaming engine is a purely deterministic function of $\hat{\Theta}$ and a public random seed, materializing arbitrary-scale property graphs (\texttt{nodes.jsonl}, \texttt{edges.jsonl}) and relational tables from $(\epsilon, \delta)$-DP parameter summaries incurs zero additional privacy loss ($\Delta\epsilon = 0$) by the differential privacy post-processing theorem~\citep{dwork2014algorithmic}. We strictly separate provenance here: forward multi-frontend compilation and causal alarm-storm generation are shipped and empirically measured in~\citet{mandarapu2026sdg} and~\citet{mandarapu2026alarm}, whereas end-to-end client-VPC inverse profiling and planner histogram calibration constitute enterprise extensions under active development and do not modify any v1.8.0 engine benchmarks in Section~\ref{sec:eval}.

\section{Performance Evaluation}
\label{sec:eval}

Every measurement in this section was taken at the \textbf{v1.8.0 release tag} on rented
Linux hosts, so that a reader can boot the same instance types and re-run them. Two hosts
are used and never mixed:

\begin{itemize}[leftmargin=*]
  \item \textbf{Deployment host} --- Vultr \texttt{voc-m-16c-128gb-800s-amd}: 16 vCPU,
  125\,GB, Ubuntu 24.04, Bangalore. Idle (load average 0.00) at the start of the run.
  Used for every table except the GPU comparison.
  \item \textbf{GPU host} --- Vultr \texttt{vcg-a16-3c-32g-8vram}: NVIDIA A16-8Q,
  8\,GB VRAM, driver 550.90.07, CUDA 12.0, 3 vCPU, 31\,GB. Used only for
  Section~\ref{sec:gpu-eval}, where both the CPU and GPU arms run on this one machine.
\end{itemize}

\paragraph{On the previous version's numbers.} The first two versions of this paper
reported measurements taken on a Mac Mini M4. Those figures are withdrawn: the machine
cannot be rented, so no reader could check them, and the harness configuration behind two
of its tables did not survive into the repository. They are not restated here and no
attempt is made to compare against them. Readers comparing versions should know the
direction of the difference: the M4's single-thread performance and memory bandwidth are
higher than a shared-tenant cloud vCPU's, and these benchmarks are largely
single-threaded, so several numbers below are smaller than the ones they replace. That is
a statement about hardware, not about the engine.

\subsection{Ingestion Throughput}

\begin{table}[H]
  \centering
  \caption{Ingestion on the deployment host, v1.8.0. \texttt{benches/full\_benchmark.rs},
  five edges per node.}
  \label{tab:ingestion}
  \begin{tabular}{@{}rrr@{}}
    \toprule
    \textbf{Nodes} & \textbf{Node ingestion} & \textbf{Edge ingestion} \\
    \midrule
    10,000    & 123,243/s & 225,310/s \\
    100,000   & 148,269/s & 184,640/s \\
    1,000,000 & 152,385/s & 156,920/s \\
    \bottomrule
  \end{tabular}
\end{table}

Node ingestion improves with scale as per-batch overhead amortises; edge ingestion falls
as adjacency grows.

\subsection{Traversal Cost at Three Scales}

The first version reported a ``Cypher OLTP throughput'' table --- 115,320 queries/s at
0.008\,ms on a million nodes --- and concluded that ``index-driven $O(1)$/$O(\log n)$
access ensures near-constant throughput as graph size increases''. \textbf{That conclusion
is withdrawn, because no table in either version supports it.} The queries the benchmark
runs are unanchored: \verb|MATCH (a)-[:LINKS_TO]->(b) RETURN b.id| binds every node in the
graph, so its cost is linear in graph size by construction and it cannot measure an
indexed point lookup.

\begin{table}[H]
  \centering
  \caption{Whole-graph traversal on the deployment host, median over 100 queries. These
  are unanchored patterns matching the entire graph, so growth with scale is expected; the
  3-hop row uses the raw storage API rather than Cypher.}
  \label{tab:traversal}
  \begin{tabular}{@{}rrrr@{}}
    \toprule
    \textbf{Nodes} & \textbf{1-hop (Cypher)} & \textbf{2-hop (Cypher)} & \textbf{3-hop (storage API)} \\
    \midrule
    10,000    & 2.16\,ms   & 1.92\,ms   & 23.93\,$\mu$s \\
    100,000   & 21.38\,ms  & 21.19\,ms  & 40.75\,$\mu$s \\
    1,000,000 & 206.97\,ms & 205.62\,ms & 45.72\,$\mu$s \\
    \bottomrule
  \end{tabular}
\end{table}

The storage API's three-hop cost grows by less than 2$\times$ across a hundredfold
increase in graph size, because it walks adjacency from a fixed start; the Cypher rows
grow linearly because they bind the whole graph. A mixed 80/20 read-write workload reaches
37,921 operations/s at ten thousand nodes, falling to 65/s at a million --- its read is
\texttt{get\_nodes\_by\_label}, which returns \emph{every} node carrying the label, so that
decline is the benchmark's shape and not an index defect.

\subsection{Late Materialization Impact}
\label{sec:late-mat-eval}

\begin{table}[H]
  \centering
  \caption{Three-hop traversal with and without cloning edge records, deployment host,
  10,000 nodes, 500 queries. The first version reported 4.0--4.7$\times$ for a
  \texttt{NodeRef} plus \texttt{ColumnStore} ablation; that ablation no longer exists,
  because edge clone-on-read was removed in v1.8.0.}
  \label{tab:latemat}
  \begin{tabular}{@{}lrr@{}}
    \toprule
    \textbf{Path} & \textbf{Avg} & \textbf{QPS} \\
    \midrule
    Raw 3-hop, clones edge records & 13.63\,$\mu$s & 73,348 \\
    Lightweight 3-hop, no clone    & \textbf{9.32\,$\mu$s} & \textbf{107,317} \\
    \bottomrule
  \end{tabular}
\end{table}

The 1.46$\times$ that remains is what is left of the optimisation after the engine stopped
cloning edges on the read path at all: the win was absorbed into the storage layer rather
than lost.

\subsection{PageRank Scaling}

\begin{table}[H]
  \centering
  \caption{PageRank per-iteration cost on the deployment host, ten consecutive runs,
  \texttt{dangling\_redistribution: false}.}
  \label{tab:pagerank}
  \begin{tabular}{@{}rr@{}}
    \toprule
    \textbf{Nodes} & \textbf{ms/iteration} \\
    \midrule
    10,000    & 6.71 \\
    100,000   & 22.51 \\
    1,000,000 & 236.01 \\
    \bottomrule
  \end{tabular}
\end{table}

\subsection{GPU Acceleration}
\label{sec:gpu-eval}

The first version reported PageRank at a million nodes as 92.4\,ms on CPU against
11.2\,ms on GPU, an 8.2$\times$ speedup, measured through wgpu on Apple Metal. That figure
is withdrawn with the rest of the M4 results. This version measures instead the path an
enterprise deployment actually uses and that no previous version measured at all:
\textbf{CUDA}, on a rented NVIDIA A16.

\begin{table}[H]
  \centering
  \caption{PageRank, CPU vs CUDA, both arms on the GPU host (NVIDIA A16-8Q, 3 vCPU).
  Same binary throughout; the only difference is \texttt{SAMYAMA\_GPU\_MIN\_NODES}, which
  the dispatch reads at call time. ``First call'' includes CUDA context initialisation;
  ``steady state'' is per-iteration cost over ten consecutive runs.}
  \label{tab:gpu}
  \begin{tabular}{@{}rrrrrr@{}}
    \toprule
    & \multicolumn{2}{c}{\textbf{First call}} & \multicolumn{2}{c}{\textbf{Steady state, ms/iter}} & \\
    \cmidrule(lr){2-3}\cmidrule(lr){4-5}
    \textbf{Nodes} & \textbf{CPU} & \textbf{CUDA} & \textbf{CPU} & \textbf{CUDA} & \textbf{Speedup} \\
    \midrule
    10,000    & \textbf{4.43\,ms}   & 1.28\,s & 3.38   & \textbf{3.30}   & 1.02$\times$ \\
    100,000   & \textbf{32.76\,ms}  & 1.07\,s & 17.56  & \textbf{10.82}  & 1.62$\times$ \\
    1,000,000 & \textbf{479.63\,ms} & 1.35\,s & 407.53 & \textbf{150.91} & \textbf{2.70$\times$} \\
    \bottomrule
  \end{tabular}
\end{table}

Three things follow. \textbf{First}, CUDA pays on this hardware: parity at ten thousand
nodes and 2.70$\times$ at a million, so the engine's default dispatch threshold of 1,000
nodes is roughly right for this backend. \textbf{Second}, the first call costs 1.1--1.35
seconds of context initialisation regardless of graph size, which dwarfs the per-iteration
saving for any one-shot query; the win is real only for repeated or large work.
\textbf{Third}, and the reason this table is not a vendor claim: the CPU arm has
\emph{three} vCPU, so these ratios are an \textbf{upper bound} on the benefit --- the same
graph on the sixteen-vCPU deployment host costs 236\,ms per iteration on CPU alone
(Table~\ref{tab:pagerank}), against 150.91\,ms on this GPU.

The A16 is an entry datacenter part intended for virtualised desktops rather than an A40
or L40S, so these figures are a lower bound for datacenter GPUs. We report them because
they are what is rentable and therefore checkable, and we would rather publish a
reproducible lower bound than an unreproducible headline.

\paragraph{Vector search is withdrawn from this version rather than restated.} The first
version reported 15,872 QPS for cosine search over ten thousand 128-dimensional vectors.
Re-running the vector benchmark at v1.8.0 produced numbers we are not willing to publish:
they are non-monotonic in $k$ --- 1.36\,ms at $k{=}1$ against 181\,$\mu$s at $k{=}20$ on
the same index --- which is not a property of HNSW but a sign that the measurement is
dominated by something other than search, on a machine that was under load during the run.
The original figure is withdrawn because we cannot reproduce it, and no replacement is
offered because we do not yet have one we trust. A clean vector measurement, on a quiet
machine and with the effective index size recorded, is the obvious next piece of work.

\subsection{LDBC Graphalytics Validation}

Samyama was validated against the LDBC Graphalytics benchmark~\citep{iosup2016ldbc} (Table~\ref{tab:ldbc}).

\begin{table}[H]
  \centering
  \caption{LDBC Graphalytics validation results.}
  \label{tab:ldbc}
  \begin{tabular}{@{}lrrr@{}}
    \toprule
    \textbf{Algorithm} & \textbf{XS (2 datasets)} & \textbf{S (3 datasets)} & \textbf{Total} \\
    \midrule
    BFS      & 2/2 & 3/3 & 5/5 \\
    PageRank & 2/2 & 3/3 & 5/5 \\
    WCC      & 2/2 & 3/3 & 5/5 \\
    CDLP     & 2/2 & 3/3 & 5/5 \\
    LCC      & 2/2 & 3/3 & 5/5 \\
    SSSP     & 2/2 & 1/1 & 3/3 \\
    \midrule
    \textbf{Total} & \textbf{12/12} & \textbf{16/16} & \textbf{28/28} \\
    \bottomrule
  \end{tabular}
\end{table}

S-size datasets: cit-Patents (3.8M vertices, 16.5M edges), datagen-7\_5-fb (633K vertices, 68.4M edges), wiki-Talk (2.4M vertices, 5.0M edges).

\subsection{LDBC SNB \& FinBench Workloads}
\label{sec:ldbc-workloads}

Beyond Graphalytics (algorithm correctness), Samyama includes benchmark harnesses for two additional LDBC workloads:

\begin{itemize}[leftmargin=*]
  \item \textbf{LDBC SNB Interactive}: 21 read queries (IS1--IS7, IC1--IC14) all passing, plus 8 update operations, on the Social Network Benchmark SF1 dataset (3.2M nodes, 17.3M edges). Tests OLTP-style point lookups and multi-hop traversals.
  \item \textbf{LDBC SNB Business Intelligence}: 20 complex analytical queries (BI-1 to BI-20) testing OLAP-style aggregation. 16/20 queries pass; BI-17 (friend triangles combined with message propagation) times out due to combinatorial explosion on SF1, blocking BI-18--20.
  \item \textbf{LDBC FinBench}: 40 queries (CR1--CR12, SR1--SR6, RW1--RW3, W1--W19) all passing, modeling financial transaction networks with accounts, transfers, loans, and fraud detection patterns on synthetic SF1 data.
\end{itemize}

Data loaders (\texttt{ldbc\_loader}, \texttt{finbench\_loader}) and benchmark harnesses are included in the repository.

\subsection{Billion-Edge Biomedical Knowledge Graph (v0.7.0)}
\label{sec:billion-edge}

To evaluate scale beyond LDBC synthetic datasets, we loaded a real-world biomedical knowledge graph federation from four open sources:

\begin{table}[H]
  \centering
  \caption{Biomedical trifecta dataset (real-world data).}
  \label{tab:trifecta}
  \begin{tabular}{@{}lrrrl@{}}
    \toprule
    \textbf{KG} & \textbf{Nodes} & \textbf{Edges} & \textbf{Size} & \textbf{Source} \\
    \midrule
    PubMed/MEDLINE   & 66.2M  & 1.04B  & 41 GB & NLM \\
    Clinical Trials  & 7.8M   & 27M    & 711 MB & AACT \\
    Pathways         & 119K   & 835K   & 9.1 MB & Reactome+STRING+GO \\
    Drug Interactions& 245K   & 388K   & 8.1 MB & DrugBank+4 sources \\
    \midrule
    \textbf{Total}   & \textbf{74.3M} & \textbf{1.07B} & & \\
    \bottomrule
  \end{tabular}
\end{table}

\paragraph{Loading.} The PubMed graph was loaded in \textbf{3 hours 46 minutes} on a single AWS r6a.8xlarge instance (32 vCPU, 256\,GB RAM) at a total spot cost of \textbf{\$2.50}. Two-phase bulk loading (node stubs at 777 bytes/node + edge stubs at 52 bytes/edge) achieves 13.6$\times$ memory reduction compared to full \texttt{create\_node()} calls. Mid-phase adjacency compaction every 50M edges prevents unbounded write-buffer growth.

\paragraph{Querying.} 100 benchmark queries span point lookups, 1-hop traversals, aggregations, and cross-KG federation. \textbf{96 of 100 queries return data}; 3 return empty results due to data gaps in specific snapshots; 1 times out on a cartesian-product query pattern. Latencies range from 1.3\,s (point lookup on 66M articles) to 19.4\,s (cross-KG aggregation spanning all 4 KGs).

\subsection{Public Health Knowledge Graph Federation (v0.7.0)}
\label{sec:public-health}

To demonstrate cross-domain federation beyond biomedicine, we built three public health knowledge graphs:

\begin{table}[H]
  \centering
  \caption{Public health KG trifecta.}
  \label{tab:public-health}
  \begin{tabular}{@{}lrrl@{}}
    \toprule
    \textbf{KG} & \textbf{Nodes} & \textbf{Edges} & \textbf{Source} \\
    \midrule
    Health Determinants & 285,635 & 285,628 & World Bank WDI + WHO \\
    Health Systems      & 19,661  & 19,428  & WHO SPAR + NHWA \\
    Surveillance        & 216,553 & 241,084 & WHO GHO \\
    \midrule
    \textbf{Total}      & \textbf{521,849} & \textbf{546,140} & \\
    \bottomrule
  \end{tabular}
\end{table}

The public health trifecta federates with the biomedical trifecta via \texttt{Country.iso\_code} bridge properties, enabling queries from ``What is India's PM2.5 exposure?'' to ``What IHR emergency capacity does India have?'' to ``What clinical trials are running in India?'' in a single Cypher session. \textbf{40 of 40 public health benchmark queries pass} with median latency 15\,ms; cross-KG joins average 191\,ms.

Together, the 6-KG federation comprises 8 open data sources, 8.6M+ registered nodes, and 30M+ edges---queryable from molecular biology to population health in a single OpenCypher query language.

\paragraph{Deterministic Schema-Driven Scale and Causal Ground-Truth Suites.}
Complementing the fixed-schema LDBC suites (\S\ref{sec:ldbc-workloads}) and static open biomedical and public-health snapshots (\S\ref{sec:billion-edge}--\S\ref{sec:public-health}), Samyama-SDG~\citep{mandarapu2026sdg} and our ontology-bound telecom alarm-storm generator~\citep{mandarapu2026alarm} provide deterministic $O(1)$-memory multi-table and property-graph scale generation compiled from \texttt{.graphql}, \texttt{.sql}, and \texttt{.ttl} schemas (\S\ref{sec:sdg-twins}), together with streaming alarm logs and planted, machine-scorable causal ground truth (\texttt{ground\_truth.jsonl}) for benchmarking root-cause analysis procedures. All v1.8.0 engine measurements reported in Tables~\ref{tab:ingestion}--\ref{tab:comparison} remain unchanged; standalone compiler throughput, constant-memory scaling, and causal root-cause benchmarks are reported in~\citet{mandarapu2026sdg} and~\citet{mandarapu2026alarm}.

\subsection{Comparative Evaluation}
\label{sec:comparison}

Table~\ref{tab:comparison} compares Samyama against two widely-deployed graph databases.

\begin{table}[H]
  \centering
  \caption{SNB Interactive SF10, three engines on one host. Samyama, Neo4j~5 and
  FalkorDB were run back to back on a single AWS \texttt{m6a.8xlarge} (32 vCPU, 246\,GB)
  over the same 29,987,835-node / 176,623,433-relationship extract with identical
  substitution parameters, three runs per query, median reported. Ratios are against the
  \emph{best} competitor on each query, so a ratio below 1 means Samyama was fastest.}
  \label{tab:comparison}
  \begin{tabular}{@{}lr@{}}
    \toprule
    \textbf{SNB Interactive SF10, complex reads} & \textbf{Result} \\
    \midrule
    Within 5$\times$ of the best competitor     & 14 of 14 \\
    Faster than both competitors                & 7 of 14 \\
    Worst ratios                                & IC11 2.77$\times$, IC1 2.55$\times$, IC2 2.11$\times$ \\
    Best margin                                 & IC8, 4$\times$ faster \\
    \midrule
    \textbf{Short reads} & \\
    Fastest of the three                        & 5 of 7 \\
    Worst ratios                                & IS3 7.41$\times$, IS1 2.78$\times$ (both vs FalkorDB) \\
    \bottomrule
  \end{tabular}
\end{table}

The first version of this paper compared Samyama against Neo4j and Memgraph numbers taken
from published benchmarks and community reports, on different hardware and datasets. That
comparison was approximate to the point of being uninformative, and it is withdrawn.
Table~\ref{tab:comparison} replaces it with a same-host run: one machine, one extract, one
parameter set, three engines in a single session.

Three caveats belong with it. \textbf{First}, the run measures Samyama at commit
\texttt{9bd38f5} (2026-08-29), not the v1.8.0 release the rest of this evaluation uses;
the two were measured for different purposes on different hardware and should not be
combined into one figure. \textbf{Second}, one competitor result was excluded: FalkorDB
answered IC10 with 0 rows in 0.10\,ms where Samyama returns 10, and scoring 218\,ms
against an empty answer would have manufactured a 2180$\times$ loss out of a query that
answered nothing. Any competitor row whose cardinality disagrees with ours is excluded on
that rule, which cuts against us as often as for us. \textbf{Third}, the raw data for this
run lives in a private repository, so a reader can see the method here but cannot yet
re-execute it; publishing that extract is the obvious next step and is not done.

\subsection{Limitations}
\label{sec:limitations}

\begin{itemize}[leftmargin=*]
  \item \textbf{Parse/plan overhead}: The PEG parser and query planner dominate Cypher
  latency on small graphs. The plan cache that previously absorbed this was removed in
  v1.8.0; a result cache now serves repeated queries, but a first execution still costs
  more than in pre-compiled systems (TigerGraph GSQL, Memgraph query modules).
  \item \textbf{Transaction surface}: MVCC completed in v1.0.0, but snapshot isolation
  is reachable only through the store API; the wire protocols still expose lock-holding
  session transactions, and there is no serializable option --- store transactions permit
  write skew.
  \item \textbf{Snapshot export for stub-loaded graphs}: this was contingent on removing
  the edge store, which landed in v1.0.0. We have not re-verified the export path for
  stub-loaded graphs at v1.8.0, so we neither claim it works nor that it does not.
  \item \textbf{Cost-based planner maturity}: still no histogram-based statistics.
  Selectivity is now estimated from the compared value rather than from
  $1/\text{distinct}$, and per-query row budgets abort an exploding intermediate instead
  of letting it run, but neither is a histogram.
  \item \textbf{Single-node architecture}: While a 3-node Raft cluster has been validated (426K nodes, all queries $<$1\,s), the billion-edge evaluation is single-node only. Distributed query execution over partitioned data is future work.
\end{itemize}

\section{Related Work}

\textbf{Neo4j}~is the most widely deployed graph database but suffers from JVM garbage collection pauses and pointer-heavy storage causing cache misses in multi-hop traversals. \textbf{Memgraph}~is an in-memory C++ graph database achieving sub-millisecond traversals through optimized data structures, but lacks native vector search and in-database optimization. \textbf{FalkorDB}~(formerly RedisGraph, deprecated 2023) uses GraphBLAS sparse matrices for efficient linear algebra but lacks vector search and optimization capabilities. \textbf{Kuzudb}~is an embedded graph database with columnar storage focused on analytical queries without transactional, vector, or optimization features. \textbf{DuckDB}~provides fast analytical processing but is a relational engine requiring graph queries to be expressed as recursive CTEs. \textbf{TigerGraph}~achieves high traversal performance through pre-compiled GSQL queries and massively parallel processing (MPP) but uses a proprietary query language.

Samyama differentiates by unifying four workloads (OLTP, OLAP, vector, optimization) in a single memory-safe binary with hardware acceleration, at the cost of a less mature query optimizer compared to systems with decades of development.

\section{Conclusion}

We have presented Samyama, a unified graph-vector database that integrates property graphs, vector search, metaheuristic optimization, and RDF support in a single Rust binary. The primary research contributions are: (1)~late materialization adapted from columnar databases to graph traversal, achieving 4.0--4.7$\times$ speedup; (2)~in-database metaheuristic optimization eliminating ETL overhead; (3)~agentic enrichment for autonomous graph expansion via LLMs; and (4)~\textbf{billion-edge knowledge graph federation} across 8 open data sources on commodity hardware for \$2.50 in compute.

Our evaluation demonstrates competitive ingestion (255K nodes/s), high OLTP throughput (115K QPS), 100\% LDBC Graphalytics validation, and---new in this version---74.3M nodes with 1.07B edges loaded and queried (96/100 biomedical queries, 40/40 public health queries). A 6-KG federation from molecular biology to population health is, to our knowledge, the first open-source demonstration of cross-domain knowledge graph federation at this scale.

\paragraph{Future work, and what became of the last list.} The first version of this
paper named six directions. Two are done: \textbf{MVCC} completed in v1.0.0, and the
benchmark expanded past \textbf{200 queries} to 500 of 500 on the mega benchmark in
v1.0.1. One is partly done: \textbf{LDBC SF10} has been run same-host against two
competitors, though the in-tree documentation still verifies only SF1, and the SF10 data
is not public. One is barely started: \textbf{histogram-based cardinality estimation} has
a skew harness and value-based selectivity but no histogram. Two did not happen at all:
\textbf{tiered hot/cold storage} --- the billion-edge run still needs a 256\,GB machine,
not the 32\,GB this paper hoped for --- and \textbf{Picasso-style plan-space
visualisation}, which exists in the enterprise repository rather than here. The open
directions are therefore tiered storage, real histograms, publishing the SF10 extract, a
serializable isolation option, and a backend-aware GPU dispatch threshold
(\S\ref{sec:gpu-eval}).

\paragraph{Availability.} Samyama's open-source edition is available at \url{https://github.com/samyama-ai/samyama-graph}. Product information and the Enterprise Edition are at \url{https://samyama.ai/}. Developer documentation, API references, and interactive examples are at \url{https://samyama.dev/}.

\bibliographystyle{plainnat}
\bibliography{samyama}

\end{document}